# AI driven B-cell Immunotherapy Design

Bruna Moreira da Silva[1,2,3], David B. Ascher[1,2,4], Nicholas Geard[3], and Douglas E. V. Pires[1,3,*]

[1]Systems and Computational Biology, Bio21 Institute, University of Melbourne, Melbourne, Victoria, Australia

[2]Computational Biology and Clinical Informatics, Baker Heart and Diabetes Institute, Melbourne, Victoria, Australia

[3]School of Computing and Information Systems, University of Melbourne, Melbourne, Victoria, Australia

[4]The School of Chemistry and Molecular Biosciences, The University of Queensland, Brisbane, Queensland, Australia

*To whom correspondence should be addressed D.E.V.P. Email: douglas.pires@unimelb.edu.au.

**Bruna Moreira da Silva** is a PhD student at The University of Melbourne. Her research interests are in bioinformatics, immunoinformatics and machine learning to advance Global Health.

**David B. Ascher** is the Director of Biotechnology at the University of Queensland and head of Computational Biology and Clinical Informatics at the Baker Institute and Systems and Computational Biology at Bio21 Institute. He is interested in developing and applying computational tools to assist leveraging clinical and omics data for drug discovery and personalised medicine.

**Nicholas Geard** is an Associate Professor at the School of Computing and Information Systems at the University of Melbourne and Director of the Melbourne Data Analytics Platform. He is a computer scientist specialising in computational simulation applied to a range of problems in health and epidemiology.

**Douglas E. V. Pires** is an Associate Professor in Digital Health at the School of Computing and Information Systems at the University of Melbourne and group leader at the Bio21 Institute. He is a computer scientist and bioinformatician specialising in machine learning and AI and the development of the next generation of tools to analyse omics data, and guide drug discovery and personalised medicine.

**ABSTRACT**

Antibodies, a prominent class of approved biologics, play a crucial role in detecting foreign antigens. The effectiveness of antigen neutralisation and elimination hinges upon the strength, sensitivity, and specificity of the paratope-epitope interaction, which demands resource-intensive experimental techniques for characterisation. In recent years, artificial intelligence and machine learning methods have made significant strides, revolutionising the prediction of protein structures and their complexes. The past decade has also witnessed the evolution of computational approaches aiming to support immunotherapy design. This review focuses on the progress of machine learning-based tools and their frameworks in the domain of B-cell immunotherapy design, encompassing linear and conformational epitope prediction, paratope prediction, and antibody design. We mapped the most commonly used data sources, evaluation metrics, and method availability and thoroughly assessed their significance and limitations, discussing the main challenges ahead.

# INTRODUCTION

Therapeutic antibodies are a rapidly growing class of biopharmaceuticals with potentially exceptional antigen specificity and affinity. Their ability to detect and eliminate a wide array of foreign threats makes them suitable for a range of potential therapeutic and diagnostic applications. Antibody and antigen engineering have been greatly benefited by the evolution of research in computational biology, leading to innovative approaches in screening antibody targets, optimising their biochemical and physical properties, predicting and optimising binding affinity and understanding escape mutations [1]. Antibody therapeutics reached the milestone of around 175 drugs approved or under regulatory review by the end of 2022, targeting diverse types of diseases, such as oncologicals, autoimmunes, chronics, neurodegenerative and viral infections [2].

The sophisticated mechanisms governing antibody responses are orchestrated within the cooperative subsystem known as the Adaptive Immune System, with T- and B-cell lymphocytes serving as its main actors. When antigens are recognised by B-cell receptors, these specialised white blood cells initiate a highly specific and tailored immune response by releasing antibodies that target that specific epitope, the exposed region of the antigen recognised by the immune system [3].

The amino acid residues that compose the epitope region may be arranged in two different distributions on the antigen surface: adjacent in the primary sequence, known as linear epitopes, or adjacent in the three-dimensional (3D) structure, known as conformational epitopes [4]. It is worth noting that although conformational epitopes have been the most observed [5], they contain sequential stretches of amino acids, thereby also exhibiting characteristics of linear epitopes [6], [7].

Related to antibodies, also known as immunoglobulins (Igs), its molecular structure consists of polypeptide chains of variable and constant domains that are further divided into two heavy chains and two light chains [8]. Variations in the heavy chain of the constant domain result in five antibody isotypes: IgA, IgD, IgE, IgG, and IgM. Among these, IgG is the most prevalent in humans [9], [10]. In Figure 1, a B-Cell receptor's recognition of an antigen is depicted, followed by the release of antibodies. These antibodies bind to the antigen's specific epitope region. The Y-shaped structure, representing the common IgG isotype, portrays the Fragment antigen-binding region, composed of the variable and constant domains of the light chain and a segment of the heavy chain, forming the "arm" of the Y. Typically, both the Paratope and epitope represent a small proportion of the antibody and antigen surfaces.

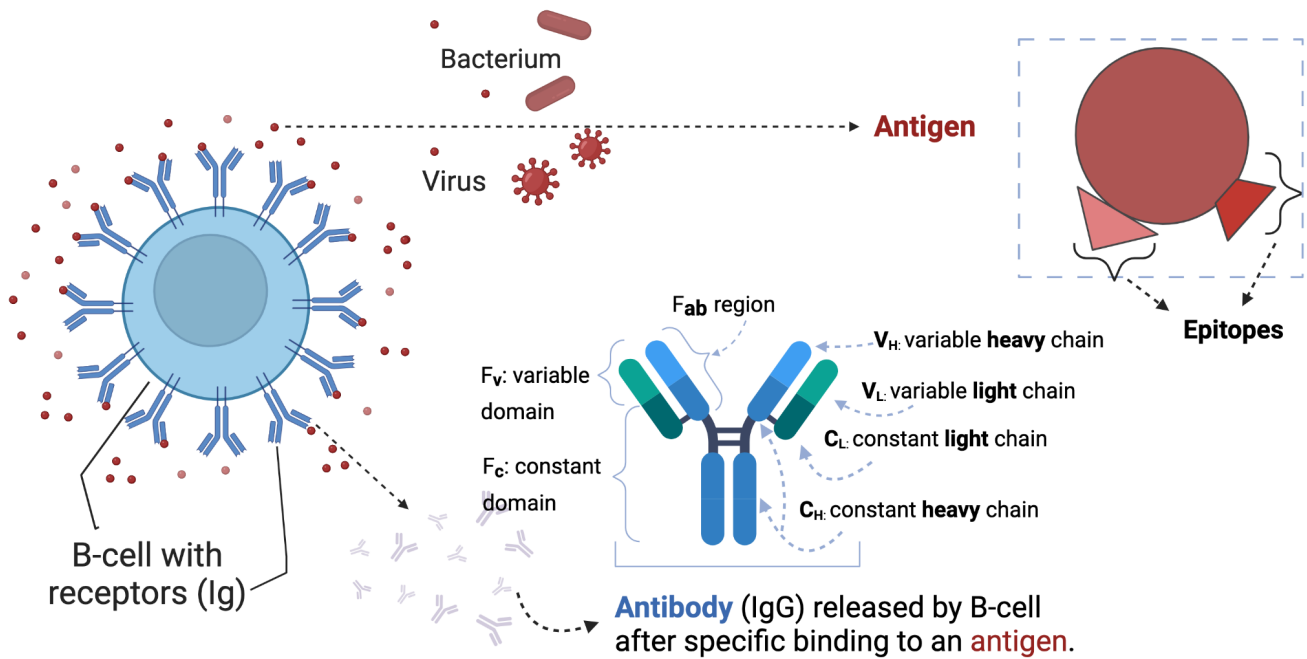

**Figure 1.** B-Cell receptors recognition of an Antigen via epitope binding, on the left side, and further production of Antibodies (IgG), in the centre, that target the specific epitope, depicted on the right. Antibody representation is basically in a Y-shape, composed of variable and constant domain chains: two heavy chains and two light chains.

Antibody specificity for various antigens is primarily achieved through differentiation in the variable regions ($F_v$). The antigen's binding site, known as the paratope, is predominantly located within these regions. They include Complementarity Determining Regions (CDR), with the third CDR in the heavy chain (CDRH3) attracting notable interest due to its broader amino acid range and sequence variability compared to other CDRs. This results in enhanced conformational specificity and diversity, enabling binding to different antigens [11]–[13].

The validation of newly identified epitopes and paratopes still requires the use of experimental techniques, for example Surface plasmon resonance (SPR) [14], [15], Nuclear magnetic resonance (NMR) [16], X-Ray Crystallography [17], cryo-EM [18], [19] and Mass Spectrometry [20]. These approaches, however, are expensive and not easily scalable, thereby limiting their applicability for comprehensive mapping of antibody-antigen (Ab-Ag) interactions.

The efforts in faster and large-scale identification of epitopes and paratopes definitely contribute to a vast range of application opportunities, especially in the antibody design field. Accordingly, these applications have evolved with the advances in biotechnology, with an unprecedented abundance of data coming from next generation sequencing (NGS) of immune repertoire and protein modelling, for instance, as with robust, refined and generalisable models in artificial intelligence (AI).

Current pipelines are composed of rounds of several sequential stages (*e.g, in vitro, in vivo* and computational approaches) combined to design a new or optimised version of a known antibody that effectively binds to a target antigen, complying with developability criteria.

Over the past decades, *in silico* tools, primarily leveraging machine learning techniques, have emerged as valuable assets to complement the limitations of experimental methods. These tools are designed to predict both linear and conformational B-cell epitopes targeting antibody-specific or antibody-agnostic regions, as well as paratopes, with most publicly available to the community either as standalone software or web servers.

Epitope identification serves as a fundamental cornerstone for various processes encompassing immunotherapies, serodiagnosis, antibody design, and vaccine development, regardless of their specific focus [21]. Mapping or predicting epitopes is a challenging task due to the interdependence of their residues with the paratope binding site [22], [23]. Shape complementarity plays a significant role, but intrinsic characteristics, such as dynamics, exposure sites, and structure, also influence the antibody binding process [24], [25]. In addition, antigen residues not participating in the binding complex with a specific Antibody might still be epitopes on a different complex [3], [26].

Prediction tools have made significant progress, thanks to the growing availability of experimental data and advancements in machine learning, particularly in protein structure prediction, antibody modelling, and overall protein engineering approaches.

The aim of this review is to summarise and highlight the evolution and developments of machine learning based tools that have been contributing to immunotherapy research, focused on prediction of B-cell Epitopes and Paratope, Antibody Design, that were made available to the community either as online repositories or web-based platforms over the last decade. The review is organised into five main sections: (i) gathering sources of available antibody and antigen data; (ii) exploring linear epitope prediction tools; (iii) analysing conformational epitope prediction tools; (iv) evaluating paratope prediction tools; and (v) assessing antibody design tools. In addition to discussing the capabilities and contributions of these tools, we critically assess their limitations, challenges, and future directions in the respective fields.

# DATA SOURCES

Antibody-antigen data derived from various experimental analyses have been deposited in extensive open repositories, with some of these summarised in Table 1 and further explained below.

Table 1. Publicly available databases including antibody and antigen experimental characterisation.

| Database | Data | Entries | Website |
|---|---|---|---|
| Protein Data Bank (PDB) | Proteins and Nucleic Acids experimentally-determined structures | 208,844 structures | https://www.rcsb.org/ |
| Structural Antibody Database (SabDab) | Antibody structures curated/annotated from the PDB | 7,632 structures | https://opig.stats.ox.ac.uk/webapps/sabdab-sabpred/sabdab |
| Antibody Database (AbDb) | Antibody structures curated/annotated from the PDB | 5,976 structures | http://www.abybank.org/abdb/ |
| IMGT/3Dstructure-DB | Antibody structures curated/annotated from the PDB | 8,616 structures | https://www.imgt.org/3Dstructure-DB/ |
| CoV-AbDab | Antibody structures reported to specifically bind to SARS-CoV-2, SARS-CoV-1 and MERS-CoV | 12,536 structures | https://opig.stats.ox.ac.uk/webapps/covabdab/ |
| Observed Antibody Space (OAS) | Antibody sequences annotated | 1,777,462 paired sequences | https://opig.stats.ox.ac.uk/webapps/oas/ |
| Immune Epitope Database (IEDB) | Epitope data from T- and B-cells | 611,502 B-cell epitopes | https://www.iedb.org/ |

The Protein Data Bank (PDB) [27] is a comprehensive biomolecules database that includes proteins, nucleic acids, and oligosaccharides. It houses annotated atomic coordinates of over 200,000 structures in three-dimensional space, primarily obtained from X-ray crystallography, cryo-EM, and NMR experiments. These structures are presented in a standardised format known as the PDB format, which organises the biological composition of amino acid sequences (and other molecule types) with corresponding atomic coordinates grouped in chains. Although the PDB is a general repository, users should utilise advanced search options to specifically filter for antibody-antigen complexes.

To facilitate the filtering process, the Structural Antibody Database (SabDab) [28], the Antibody Database (AbDb) [29] and the IMGT/3Dstructure-DB [30] have taken on the task of regularly curating and annotating only data containing antibody structures from the PDB.

Distinctively, CoV-AbDab [31] maintains a highly informative database containing a range of antibodies that are recognized to bind to betacoronaviruses, as SARS-CoV-2, SARS-CoV-1 and MERS-CoV, derived from patents and publications, in addition to providing metadata of these studies.

The Observed Antibody Space (OAS) [32] contains sequences of antibodies' variable regions in both paired and unpaired forms. These sequences are derived from 80 distinct studies of antibody repertoire sequencing and are accompanied by relevant annotations, such as individual information (*e.g.*, male, female), antibody isotype, B-cell origin (*e.g.*, plasma, naive, spleen, or peripheral blood), species, and other pertinent details.

Conversely, the Immune Epitope Database (IEDB) [33] serves as a repository for curated published experiments focused on T- and B-cells immune responses against epitopes. These experiments encompass diverse assays and organisms, encompassing millions of epitopes, primarily in peptide form. Notably, for B-cells, the database includes *in vitro* and *in vivo* study assays that involve qualitative and quantitative assessments, such as binding studies via electron microscopy, enzyme-linked immunosorbent assay (ELISA), or SPR. Additionally, the database also covers biological activities, such as neutralisation, antibody inhibition, antigen activation or agglutination.

Experimental data on antibody-antigen recognition may be subject to biases that can influence results, analysis, and interpretation of current methods. These biases encompass various aspects, including: (i) misannotation of epitope or non-epitope residues, stemming from an incomplete understanding of antigens interacting with different antibodies; (ii) overrepresentation of certain antigen organisms or sequence regions, driven by their disease significance or demand interests in research; (iii) partial or average representation of epitope-paratope interactions due to the inherent dynamics of natural proteins and the limitations of experimental techniques in accurately resolving three-dimensional conformations. Acknowledging and accounting for these biases is crucial when working with Ab-Ag experimental data to ensure more accurate and reliable outcomes and interpretations in immunotherapies research.

# LINEAR B-CELL EPITOPE PREDICTION

Linear epitopes are consecutive segments of amino acid residues in the antigen surface, ranging from approximately 5 to 25 in length, that bind to the antibody paratope. Figure 2 illustrates an Antibody-Antigen complex structure (left side), highlighting the fundamental distinction (right side) between the Linear and the Conformational epitopes: how the amino acid residues are arranged in consecutive order or not towards the paratope.

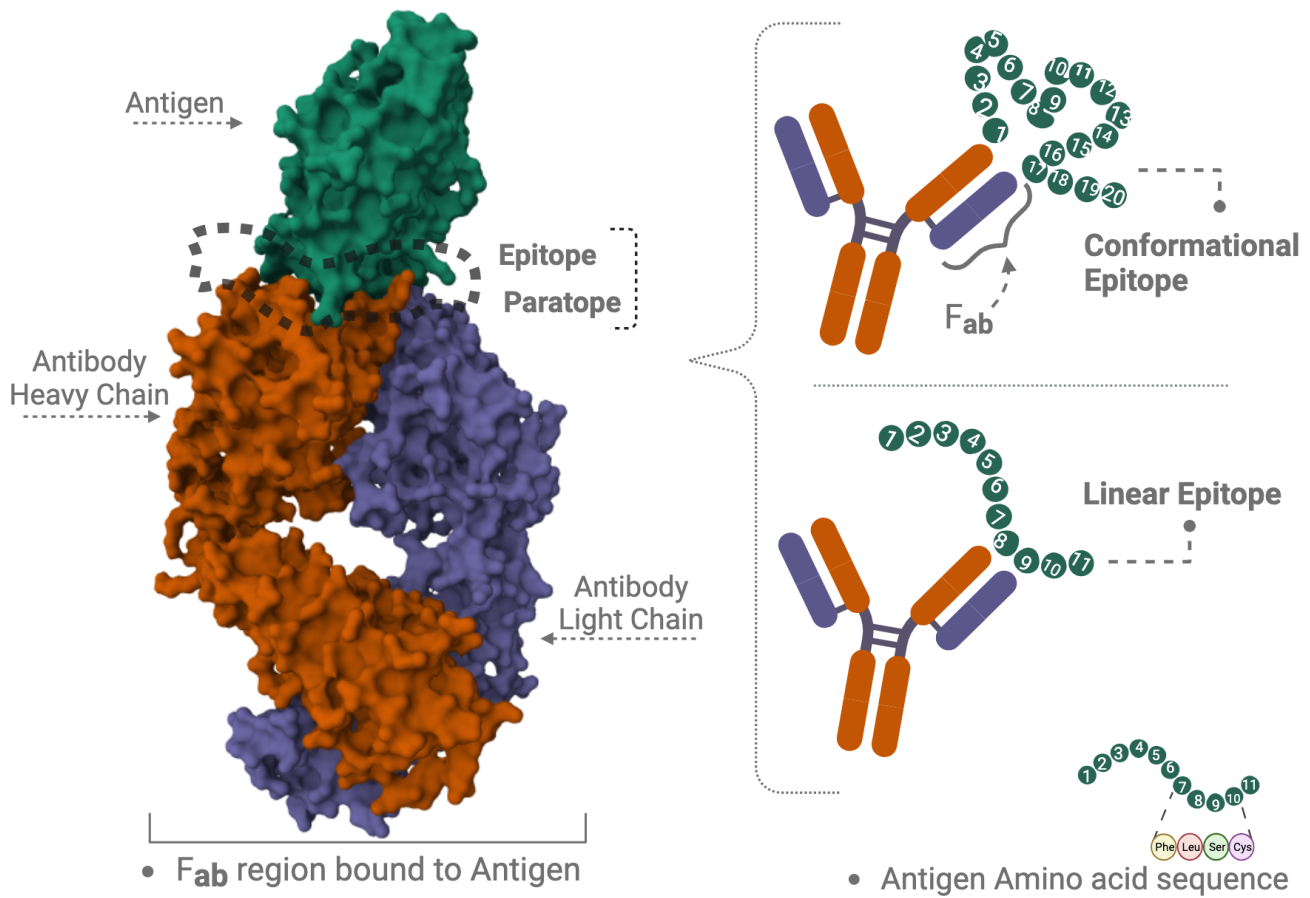

**Figure 2.** A 3D surface representation of the Antibody-Antigen complex (*PDB id:1ZTX*) on the left side. On the right side, a distinction is made between linear and conformational epitopes, related to the amino acid arrangement of the Antigen. Linear presents sequential adjacent amino acids approaching

the Antibody paratope, while Conformational displays unordered amino acid sequence, but close in space due to folding.

Most available data primarily consist of sequence information without always including the solved structure of the source antigen. Regarding epitope-paratope interaction, computational prediction of conformational epitopes is considered less frequently than linear epitopes [34]. However, predicting these epitopes through computational methods comes with inherent challenges. These challenges are particularly evident in the limited performance of existing approaches that rely on a combination of distinct amino acid encoding schemes and machine learning methods to extract patterns and characteristics from epitopes and non-epitopes. This section is divided into five subsections: (i) listing and describing Machine learning tools and their chosen algorithms for linear epitope prediction; (ii) their selected (and potentially curated) datasets; (iii) the encoding representations chosen for feature engineering; (iv) benchmarking and achieved performance; and (v) their significance, contributions and limitations.

**Machine learning-based tools**

Significant progress has been achieved in the field of linear epitope prediction, with extensive research exploring various machine learning and AI approaches. Over the past decade, multiple frameworks have been proposed and made available to the scientific community through web servers or public repositories. These frameworks accept antigen sequences as input, provided in a text-based format (*i.e.*, FASTA), and offer predicted epitope likelihood (per residue or peptide) as output. These frameworks include: BepiPred-2.0 [35], iBCE-EL [36], EpiDope [37], iLBE [38], EpitopeVEC [39], BepiPred-3.0 [40] and epitope1D [41].

Each framework utilises different machine learning architectures. Specifically, BepiPred-2.0, epitope1D, and iLBE adopted Random Forest (RF) as their chosen algorithm. iLBE further incorporated Logistic Regression (LR), while iBCE-EL employed Gradient Boosting (GB) in conjunction with Extremely Randomised Trees (ERT) and EpitopeVEC utilised Support Vector Machines (SVM). Furthermore, powered by deep neural networks, EpiDope adopted bi-directional Long Short-Term Memory (LSTM) and BepiPred-3.0 employed Feed Forward Neural Networks (FFNN).

**Datasets**

The majority of these frameworks selected the Immune Epitope Database as their data source, which provides validated experimental assays containing peptide sequences of both epitopes and non-epitopes. Given the relatively small portion of the antigen surface that epitopes occupy, a natural class imbalance in the amount of epitopes and non-epitopes is observed in databases, which may impose a burden in the machine learning process and is addressed differently by each framework.

Several of these frameworks have curated datasets to train and evaluate their machine learning models. BepiPred-2.0, iBCE-EL, EpitopeVEC, BepiPred-3.0, and epitope1D all reported using curated datasets. EpiDope and iLBE utilised datasets from BepiPred-2.0 and iBCE-EL, respectively, for comparison purposes. iBCE-EL, in particular, created a nearly balanced and non-redundant training set consisting of 9,925 peptide sequences, with 2,518 sequences used for testing. The newly released epitope1D curated the largest non-redundant set to date, preserving the natural class imbalance. It included 123,919 sequences for training and 30,980 for testing. Additionally, epitope1D retained organism information for each sequence to explore potential benefits of incorporating taxonomy-specific information of antigens. EpitopeVEC exclusively curated data from viral species, resulting in 12,892 sequences, aiming to develop a specialised predictor focused on viral antigens.

On the other hand, BepiPred-2.0 and its successor, BepiPred-3.0, used the PDB database as their primary source. They initially extracted crystal structure data of antibody-antigen complexes and subsequently annotated epitope residues based on a distance threshold criteria. However, they retained only the antigen sequences, gathering sets of 776 and 358 sequences, respectively. Additionally, their testing sets include peptide sequences of epitopes and non-epitopes curated from IEDB.

**Feature Engineering**

The diversity of modelling and feature engineering approaches adopted by each tool to represent antigen sequences in a way to facilitate distinction of epitopes and non-epitopes are highlighted in Table 2. The challenges in modelling epitopes become noticeable with the variety of combined strategies employed, with the composition of amino acid residues, including their type, frequency of occurrence, and related physicochemical attributes, representing the most extensively explored representation.

BepiPred-2.0 utilised the predicted secondary structure to categorise the shape arrangements formed in the protein backbone, commonly classified as secondary structures: alpha-helix, beta-sheets, loops, and turns. Additionally, it evaluated the relative surface accessibility (RSA), estimated volume, hydrophobicity, and polarity patterns for each amino acid residue.

In contrast, iBCE-EL computed a combination of amino acid composition with various physicochemical properties, such as hydrophobic, hydrophilic, neutral, positively or negatively charged, absolute charge, molecular weight, aliphatic index, and fraction of turn-forming residues, for each peptide sequence. iLBE adopted the position-specific scoring matrix (PSSM)[42] to measure the similarity of an amino acid sequence to a given protein database, aiming to quantify evolutionary conservation. Subsequently, it followed the encoded profile-based amino acid frequency (PKAF) approach [43], amino acid composition, and the use of AAIndex [44], a database containing numerical indices related to amino acid biochemical and physicochemical attributes.

EpiDope leveraged embeddings from language models to harness the inherent physicochemical and structural properties of proteins. It explored a context-sensitive embedding using the language model ELMo [45], previously trained, to generate continuous vectors for each residue. Similarly, the updated

BepiPred-3.0 employed pretrained transformers from evolutionary scale models to encode residues, ESM-2 [46], combined with the length of protein sequences. On the other hand, EpitopeVEC employed a context-independent language model approach in a skip-gram architecture named ProtVec [47]. It encoded each sequence along with *k-mer* representation, amino acid composition, and antigenicity scales.

epitope1D introduced a novel flexible-length graph-based representation [48] for peptide sequences. This focused on modelling physicochemical distance patterns between residues within a peptide. Additionally, it presented a customised version of the Antigenicity scale [49] and a one-hot encoding representation of organism taxonomy. Furthermore, epitope1D utilised the Composition, Transition, and Distribution (CTD)[50] to compute patterns of physicochemical and structural properties.

**Table 2.** Data encoding strategies are summarised as Features, in the second column, with the corresponding tool name listed in the first column.

| Tools | Features |
| --- | --- |
| BepiPred-2.0 | Secondary structure; RSA; Volume, Hydrophobicity; Polarity |
| iBCE-EL | Amino Acid compositions; Physicochemical properties |
| EpiDope | ELMo |
| iLBE | Amino Acid composition, PSSM, PKAF, AAIndex |
| EpitopeVEC | Amino Acid composition, Embedding through Language Model ProtVec, k-mer representation and Antigenicity Scales |
| BepiPred-3.0 | Embedding through Language Model ESM-2, Sequence length |
| epitope1D | Graph-based Signatures, Antigenicity Scale, Organism information, CTD |

**Performance**

These frameworks commonly assessed their performance through cross-validation (CV), with fold counts ranging from 5 to 10. Moreover, they conducted external validation using independent datasets. While some tools employed a limited set of statistical metrics to gauge their machine learning models' efficiency, others employed a broader array of metrics, including those traditionally used in classification tasks: Area under the ROC Curve (ROC-AUC), Accuracy (ACC), Sensitivity (Sn), Specificity (Sp), F1 score (F1), and Matthews Correlation Coefficient (MCC). These metrics derive from components of the Confusion matrix, aiding in determining the correctness of predicted values for positive or negative classes based on ground truth (True positives, True Negatives) or errors (False positives or False negatives). By considering various aspects of the confusion matrix, this combination of metrics provides enhanced confidence and prevents potentially misleading assessments.

Table 3 summarises the type of algorithm, internal and external validation strategies, metrics, dataset sizes, and performance, as reported in their respective original papers. It is evident that a lack of standardisation in metrics poses a significant challenge for adequate performance comparisons in this field. A perspective on the reach of these tools within a testing environment is possible, yet reasonable, using the shared metric amongst them, ROC-AUC, as presented in Figure 3 - Panel A, showcasing their reported performance across different linear testing datasets, which sizes are illustrated in Panel B.

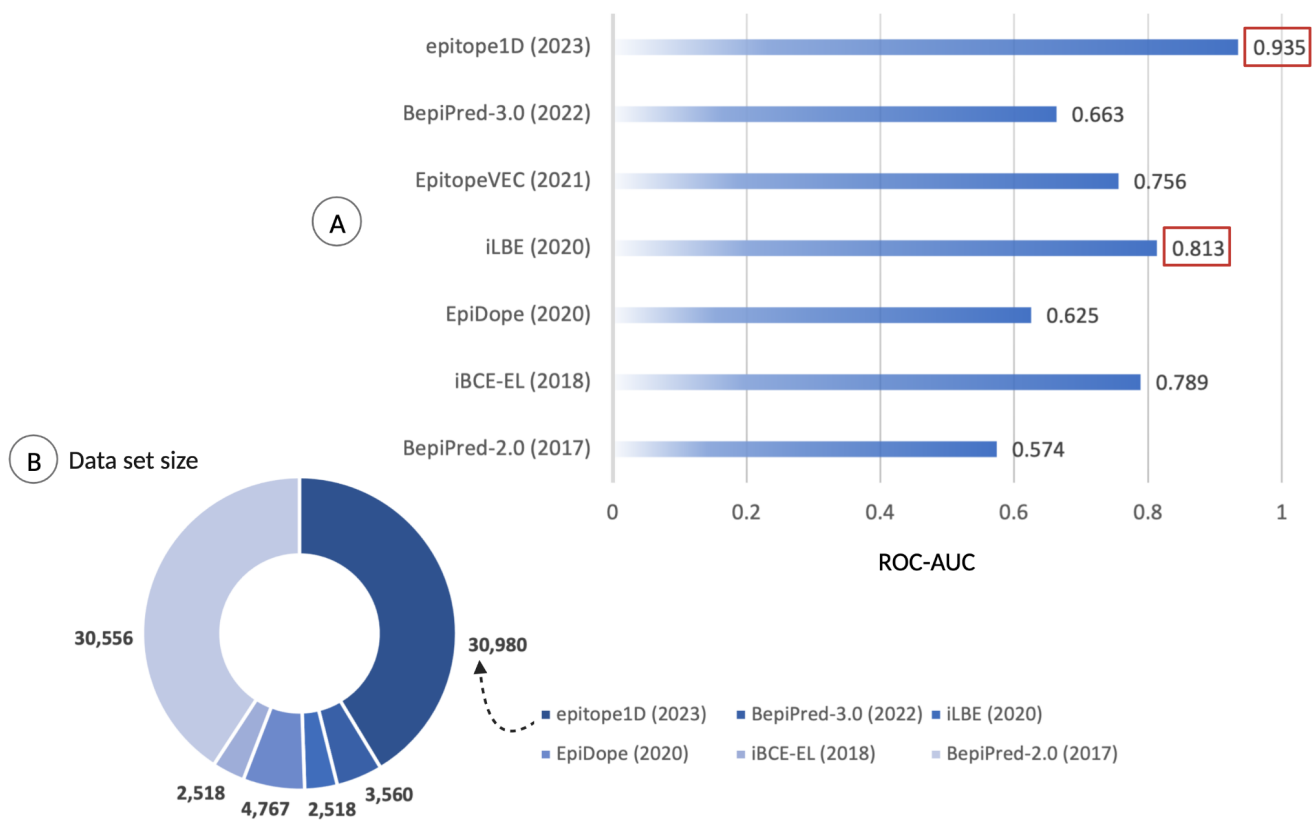

**Figure 3.** Visualisation of tool performances and data size. Panel A, on top, depicts the ROC-AUC performance of tools during blind-tests, ordered by publication year on the y-axis. Highlighted in a red square are the two highest values that exceed 0.8. Panel B, on the bottom, illustrates the sizes of their respective linear testing datasets with the dashed arrow pointing to the largest set. EpitopeVec is not shown in Panel B due to the absence of data size information.

This issue is further compounded by the use of metrics such as AUC, ACC, Sn, or Sp, as seen in BepiPred-2.0, EpiDope, and BepiPred-3.0. These metrics in isolation fail to adequately consider the varying importance of correct and incorrect predictions for the positive (epitopes) and negative (non-epitopes) classes, especially when dealing with imbalanced sets. Conversely, this constraint can be effectively addressed by adopting more adequate, balanced metrics, such as MCC [51] or F1-score, as demonstrated by iBCE-EL, iLBE, EpitopeVEC, and epitope1D. Adopting such metrics would enable more robust comparisons and evaluations of the models in this domain.

Table 3. Comparative analysis of linear B-Cell epitope predictors chronologically ordered by publication, followed by the selected machine learning algorithm, the strategies adopted to validate the model, statistical metrics employed, size of data sets and results from both cross-validation and blind testing with an independent dataset. [a]EpitopeVEC final model named Viral, tested using a viral subset version of the Bcipep database. [b]BepiPred-3.0 final model on external test set derived from the IEDB.

| Tools | Algorithm | Validation | Metrics | Dataset Size | Results on CV | Result on Independent Set |
|---|---|---|---|---|---|---|
| BepiPred-2.0 (2017) | RF | 5-fold CV using training set (PDB); Blind-test with Independent set (IEDB) | ROC-AUC, AUC10% | Training: 155 PDB ids; Testing: (1) 5 PDB ids; (2) Linear 30,556 sequences (redundant) | ROC-AUC 0.62; AUC10% 0.121 | (1) ROC-AUC 0.596; AUC10% 0.080; (2) ROC-AUC 0.574; AUC10% 0.074 |
| iBCE-EL (2018) | ERT + GB | 5-fold CV using training set (IEDB); Independent set (IEDB) | MCC, ACC, ROC-AUC, Sn, Sp | Training: 9,925 sequences; Testing: 2,518 sequences | MCC 0.454; ACC 0.729, ROC-AUC 0.782; Sn 0.716; Sp 0.739 | MCC 0.463; ACC 0.732, ROC-AUC 0.789; Sn 0.742; Sp 0.724 |
| EpiDope (2020) | Bi-directional LSTM | 10-fold CV for training set (IEDB); Independent set (IEDB) | ROC-AUC, AUC10% | Training: 24,610 sequences; Testing: 4,767 sequences | ROC-AUC 0.670; AUC10% 0.151 | ROC-AUC 0.625; AUC10% 0.120 |
| iLBE (2020) | RF + LR | 10-fold CV for training set (IEDB); | Sp, Sn, ACC, MCC, | Training: 9,925 sequences; | Sp 0.747; Sn 0.759; ACC | MCC 0.494; ROC-AUC 0.813; Sp |

| | | Independent set (IEDB) | ROC-AUC | Testing: 2,518 sequences | 0.752; MCC 0.496; ROC-AUC 0.809 | 0.745; Sn 0.752; ACC 0.748; |
|---|---|---|---|---|---|---|
| EpitopeVEC (2021) | SVM | 5-fold CV for training set[a] (IEDB); Independent set[a] (Bcipep) | ACC, Precision, Sn, F1 score, MCC, ROC-AUC | Training: 12,892 viral sequences; Testing: not informed | ACC 0.797; ROC-AUC 0.875; F1 0.850; MCC 0.554 | ACC 0.720; ROC-AUC 0.756; F1 0.541; MCC 0.264 |
| BepiPred-3.0 (2022) | FFNN | 5-fold CV for training set (PDB); Independent set[b] (IEDB) | ROC-AUC, AUC10 | Training: 343 antigens (BP3C50ID); Testing: (1) 5 structures; (2) 15 structures; (3) 3,560 linear sequences | ROC-AUC 0.762 | (1) ROC-AUC 0.738; AUC10 0.165; (2) ROC-AUC 0.771; AUC10 0.196; MCC 0.332 (3) ROC-AUC 0.663; AUC10 0.133 |
| epitope1D (2023) | RF | 10-fold CV for training set (IEDB); Independent set (IEDB) | AUC, MCC, F1 | Training: 123,919 sequences; Testing: 30,980 sequences | MCC 0.613; ROC-AUC 0.935; F1 0.658 | MCC 0.608; ROC-AUC 0.935; F1 0.654 |

Note: EPMLR (2014) [52] and LBCEPred (2022) [53] weren't included in the analysis because both web servers were down during the analysis time (February to June/2023). BepiPred-2.0 and BepiPred-3.0 were designed to address both linear and conformational epitopes.

## Significance and Limitations

The continuous improvements in performance of linear epitope prediction methods can be attributed to a combination of significant encoding representations, progressively larger non-redundant datasets, and appropriate machine learning approaches. Notably, the prediction of B-cell linear epitopes, although intrinsically challenging, has benefited considerably from quality data curation and the incorporation of new features that enhance domain knowledge with adequate abstractions.

While the use of deeper machine learning networks, such as large language models for encoding or deep neural networks as classifiers, shows promise, it is not yet definitive in significantly increasing the effectiveness of distinguishing epitopes from non-epitopes. This observation becomes particularly evident when considering the results on an independent set, as shown in the last column of Table 3. Although these results do not pertain to the same dataset, they all originate from the same source (except for EpitopeVEC). Gradual progress is observed in iBCE-EL, iLBE, and epitope1D using the ROC-AUC as a common metric, which is further corroborated by the MCC values.

# CONFORMATIONAL B-CELL EPITOPE PREDICTION

Conformational epitopes are discontinuous stretches of amino acid residues, each forming different regions on the antigen surface, and participating in binding with one or multiple antibodies [34]. In contrast to linear epitopes, the structural nature of conformational epitopes requires experimental approaches capable of capturing their 3D arrangement. Consequently, available data on conformational epitopes involves solved antigen structures at atomic level, as extensively deposited in the PDB database. As the previous section, this is similarly structured into five parts, from the machine learning approaches and datasets used, to featurisation, performance and their significance and limitations.

**Machine learning-based tools**

Frameworks developed for predicting conformational epitopes are available online as web-based platforms, similar to those presented in the linear epitope prediction section. However, due to the nature of the data involving protein structures, these approaches take the antigen structure, typically in the PDB format, as input and return the predicted probability of being an epitope (per residue).

Accordingly, SEPPA 3.0 (2019)[54], ScanNet (2022)[55], epitope3D (2022)[56], and DiscoTope-3.0 (2023)[57] adopted this approach, with SEMA (2022)[58] additionally offering a sequence-based alternative. On the other hand, SeRenDIP-CE (2021)[59] exclusively relies on a sequence-based methodology. These tools employ various machine learning algorithms. For instance, SEPPA 3.0 utilises Logistic Regression, epitope3D uses Adaptive Boosting (Adaboost), DiscoTope-3.0 employs Extreme Gradient Boosting (XGBoost), and SeRenDIP-CE relies on Random Forest. ScanNet adopts a Geometric Deep Learning approach using neural networks, while SEMA combines Transformers with a fully connected Linear layer.

**Datasets**

Solved structures of antibody-antigen complexes remain the primary evidence for conformational epitopes, making the PDB and SAbDab databases the most frequently used resources for exploring available data. To train and evaluate machine learning models, most tools manually curated the data, starting with the acquisition of antibody-antigen complexes in the PDB format. The data refinement process followed until the final antigen structures were consolidated. Some tools enumerated the datasets based on the resulting number of PDB IDs, while others considered the number of chains in each PDB ID.

Since epitope residues are not annotated in these databases, their identification becomes a subsequent step. Often, epitopes are computationally defined as antigen residues in which their heavy atoms are located within a maximum distance of 4 or 5 angstroms (Å) from an antibody. However, SEPPA 3.0 exceptionally annotated epitopes based on the calculated accessible area per residue, checking for a decrease when transitioning from the unbound state (antigen only) to the bound state (antibody-antigen complex).

From the SAbDab database, SeRenDIP-CE and ScanNet utilised a total of 280 and 796 antigen chains, respectively, which were subsequently divided for training and evaluation. DiscoTope-3.0 retrieved 24 antigens (PDB IDs) for testing only, as it used data from a previous tool, BepiPred-3.0, to train the model. From the PDB database, SEMA curated 884 antigen chains, while SEPPA 3.0 primarily focused on glycoprotein antigens, resulting in 897 chains. Additionally, epitope3D started with 1,351 PDB IDs of bound antibody-antigen complexes, later aggregating epitopes in 245 unbound antigens through structural and sequential alignments to reduce false-negative epitope annotations. Moreover, a redundancy check to decrease the sequence similarity of antigen proteins within datasets was commonly applied, except for SEPPA 3.0, which did not mention it.

**Feature Engineering**

Much like linear epitopes, encoding methods used for conformational epitopes also rely on protein-related biochemical knowledge and network-based abstractions. Tools have undergone several strategies that are summarised in Table 4.

SEPPA 3.0 utilised triangle shapes to represent amino acid residues located on the antigen surface, grouping and quantifying sets of three residues at a certain distance apart. It then identified triangles containing Asparagine glycosylation and compared their frequency in relation to epitope presence.

SeRenDIP-CE employed NetSurf [60] to predict the exposed area of amino acid residues, as measured by accessible and relative solvent accessibility (ASA and RSA). This prediction also encompassed the antigen's secondary structure, while accounting for entropy and backbone flexibility.

ScanNet adopted the concept of point clouds to iteratively cluster triplets of residues and atoms based on proximity, creating arrays that included atomic coordinates, sequence position, and residue or atom type. This representation subsequently underwent analysis via a deep geometric network.

SEMA and DiscoTope-3.0 employed a recently developed transformer language model, ESM-IF1 [61], to represent protein structures using the inverse folding approach. SEMA also utilised the ESM-1v [62] model for sequence-based inputs and included the predicted local distance difference test (pLDDT) derived from AlphaFold2 [63]. The pLDDT provides a quality score per-residue, indicating how confident the AlphaFold2 predicted structure is compared to the original structure.

Apart from utilising RSA and AAIndex, epitope3D introduced two novel features to examine the influence of epitope surroundings at both atom and residue levels. The first feature encompassed an atomic graph-based representation [64] at varying distance thresholds, incorporating physicochemical attributes. The second feature comprised a radius scanning matrix containing metrics related to residue composition within an incremental space.

**Table 4.** Final set of encoding techniques applied in each tool are presented as Features, alongside the respective tool names listed in the first column, ordered by release time.

| Tools | Features |
|---|---|
| SEPPA 3.0 | Ratio of glycosylation triangles, AAIndex. |
| SeRenDIP-CE | Accessible and Relative solvent accessibility (ASA and RSA), Secondary structure, Entropy, Flexibility. |
| ScanNet | Point clouds for atoms and amino acids |
| SEMA | ESM-IF1 and ESM-1v |
| epitope3D | Graph-based signatures, Radius Scanning Matrix, RSA, AAIndex |
| DiscoTope-3.0 | ESM-IF1, RSA, antigen length, one-hot-encoding, pLDDT |

**Performance**

Validation strategies for assessing machine learning algorithm performance in this section predominantly involved internal validation through various cross-validation schemes. This approach was utilised by SEPPA 3.0, SeRenDIP-CE, ScanNet, and epitope3D, while not being considered in SEMA and DiscoTope-3.0. A comprehensive summary of the selected algorithms, validation techniques, statistical metrics, training and test dataset sizes, and results for each tool is presented in Table 5.

When compared to predictors for linear epitopes, conformational predictors exhibit lower performance, particularly evident in terms of MCC and F1 values. To ensure an unbiased comparison across tools, the standardisation of benchmark datasets and the incorporation of appropriate metrics that address inherent class imbalance, such as MCC and PR-AUC (area under the precision recall curve), are crucial. Metrics discrepancy can be particularly exemplified in the results of the most recent tool, DiscoTope-3.0, as demonstrated in the final column of Table 5. Despite both metrics having the same range [0, 1], the PR-AUC reached 0.232, while the ROC-AUC metric achieved a value three times higher at 0.783. Hence, relying solely on metrics like ROC-AUC, Precision (same as positive predictive value, PPV), ACC could potentially lead to an overestimation of performance when faced with imbalanced classes.

**Table 5.** Comparative analysis of conformational B-Cell epitope predictors detailing the chosen algorithm, the validation strategy adopted (cross-validation, blind-test), list of statistical metrics, size of data sets (training and testing), results under cross-validation and independent set (blind-test),

| Tools | Algorithm | Validation | Metrics | Dataset Size | Results on CV | Result on Independent Set |
|---|---|---|---|---|---|---|
| SEPPA 3.0 | Logistic Regression | 10-fold CV using training set; Blind-test with Independent sets | ROC-AUC, BACC | Training: 767 chains; Testing (1): 106 chains; (2) 24 PDB ids | ROC-AUC 0.79 | (1) ROC-AUC 0.740; (2) ROC-AUC 0.749 and BACC 0.665 |
| SeRenDIP-CE | RF | 10-fold CV using training set; Blind-test with Independent sets | Precision, Recall, F1, Specificity, ACC, BACC, ROC-AUC | Training: 280 PDB ids; Testing: 56 PDB ids | Not informed | ACC: 0.684; F1: 0.259; ROC-AUC: 0.704; BACC: 0.645 |
| ScanNet | Geometric Deep Learning | 5-fold CV; Blind-test with some PDB ids | PR-AUC; PPV | Training: 796 clustered chains; Testing: 11 PDB ids | PR-AUC 0.178; PPV 0.273 | PR-AUC 0.177 |
| SEMA | ESM-1v; ESM-IF1; | Blind-test with | ROC-AUC, MCC, | Training: 884 | N/A | SEMA-1D: MCC 0.258; |

| | Linear layer. | Independent set | PPV, Sn | sequences; Testing: 101 sequences (86 PDB ids) | | AUC 0.714; PPV 0.774; SEMA-3D: MCC 0.269; ROC-AUC 0.733; PPV 0.778 |
|---|---|---|---|---|---|---|
| epitope3D | Adaboost | 10-fold CV using training set; Blind-test with Independent sets | ROC-AUC, MCC, F1, BACC | Training: 180 PDB ids; Testing: (1) 20 PDB ids; (2) 45 PDB ids | MCC 0.55; F1 0.57; BACC 0.70; ROC-AUC 0.78 | (1) MCC 0.35; F1 0.30; BACC 0.59; ROC-AUC 0.59 (2) MCC 0.45; F1 0.36; BACC 0.61; ROC-AUC 0.63 |
| DiscoTope-3.0 | XGBOOST | Blind-test with Independent sets | ROC-AUC, PR-AUC | Training: 1125 chains; Testing: (1) 281 chains; (2) 24 antigens | N/A | (1) ROC-AUC 0.807; (2) ROC-AUC 0.783; PR-AUC 0.232 |

Analogous to the previous section on linear epitopes, an illustration of tools performance in their blind tests is depicted in Figure 4. Panel A displays their model's performance using the ROC-AUC, since it's the only common metric amongst all, while Panel B showcases the corresponding dataset sizes. This reinforces the importance of employing a consistent benchmark dataset and robust statistical metrics capable of capturing class imbalance effects. Such an approach is crucial for precisely evaluating the comparative efficacy of different tools.

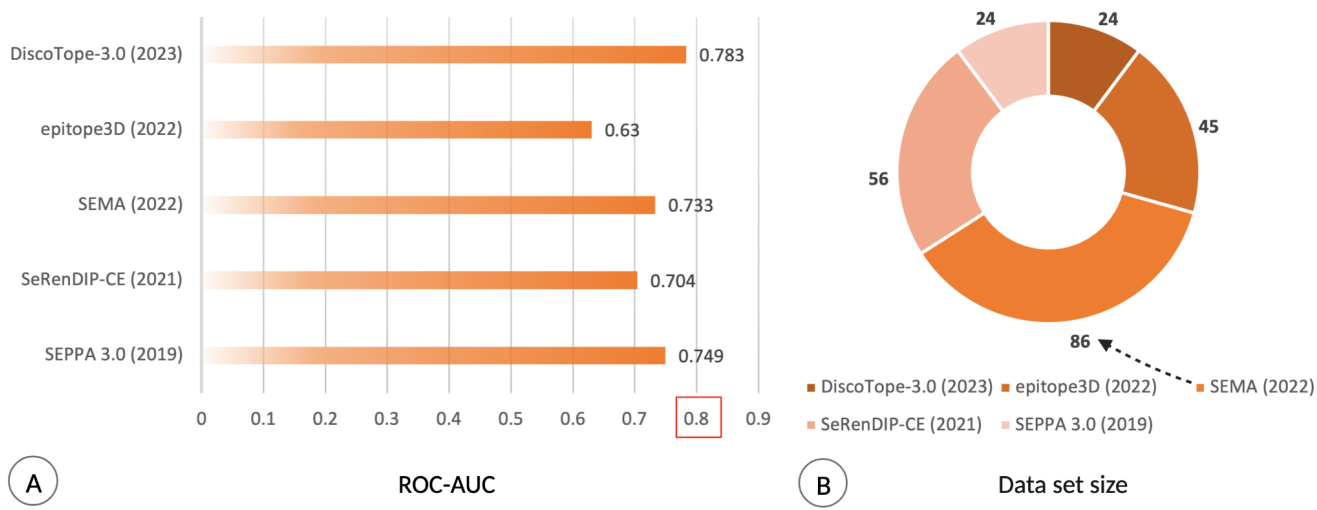

**Figure 4.** Performance illustration of the tools: In Panel A, the ROC-AUC performance during blind-tests is displayed, arranged by publication year on the y-axis. A red square on the x-axis indicates that no tool exceeds 0.8. Panel B presents the sizes of their corresponding structural testing datasets, with a dashed arrow indicating the largest dataset. ScanNet is omitted due to the absence of a shared metric with all tools.

**Significance and Limitations**

The prediction of conformational B-cell epitopes has undergone significant development through various initiatives. The process of curating high-quality dataset, with a focus on antigen representativeness, has greatly benefited from extensive and regularly updated repositories. In addition, advancements in computational techniques and the availability of easy-to-use bioinformatics tools for data preprocessing, modelling and learning have contributed to this progress. However, it's worth noting the predominance of viral antigen organisms within repositories, likely attributed to the increased demand and disease-related significance of solving antibody-virus antigen structures in experimental studies and increased pandemic preparedness.

Additionally, although the achievements in solving protein structures experimentally are expressive, as seen in the PDB database statistics mainly by X-ray crystallography method, this represents a small fraction of the vast universe of antibody and antigen conformations. Efforts employing cryoEM can potentially represent an increase in solving protein structures due to the ease of working with non crystallised samples, which is a labour intensive process and limits large and flexible molecules.

Computational modelling antigen structures has involved a spectrum of approaches over the years, ranging from established measures of exposed amino acid areas and physicochemical attributes, graph abstractions and learned embeddings using Large Language Models. Furthermore, the assessment and choice of appropriate algorithms, particularly within supervised learning contexts, have aligned with trends in the machine learning field. These trends have been harnessed through the utilisation of ensemble methods (bagging or boosting) and Deep Neural Networks. Despite these advancements, it's important to acknowledge that the performance of these tools still falls short of the ideal. This highlights the potential for enhancements and further exploration in this domain.

# PARATOPE PREDICTION

From the antibody viewpoint, the constitutive regions and amino acid sequences are more comprehensively characterised. The paratope region consists of amino acid residues within the Fragment antigen-binding region (Fab), primarily situated in the Variable regions, as illustrated in Figure 5.

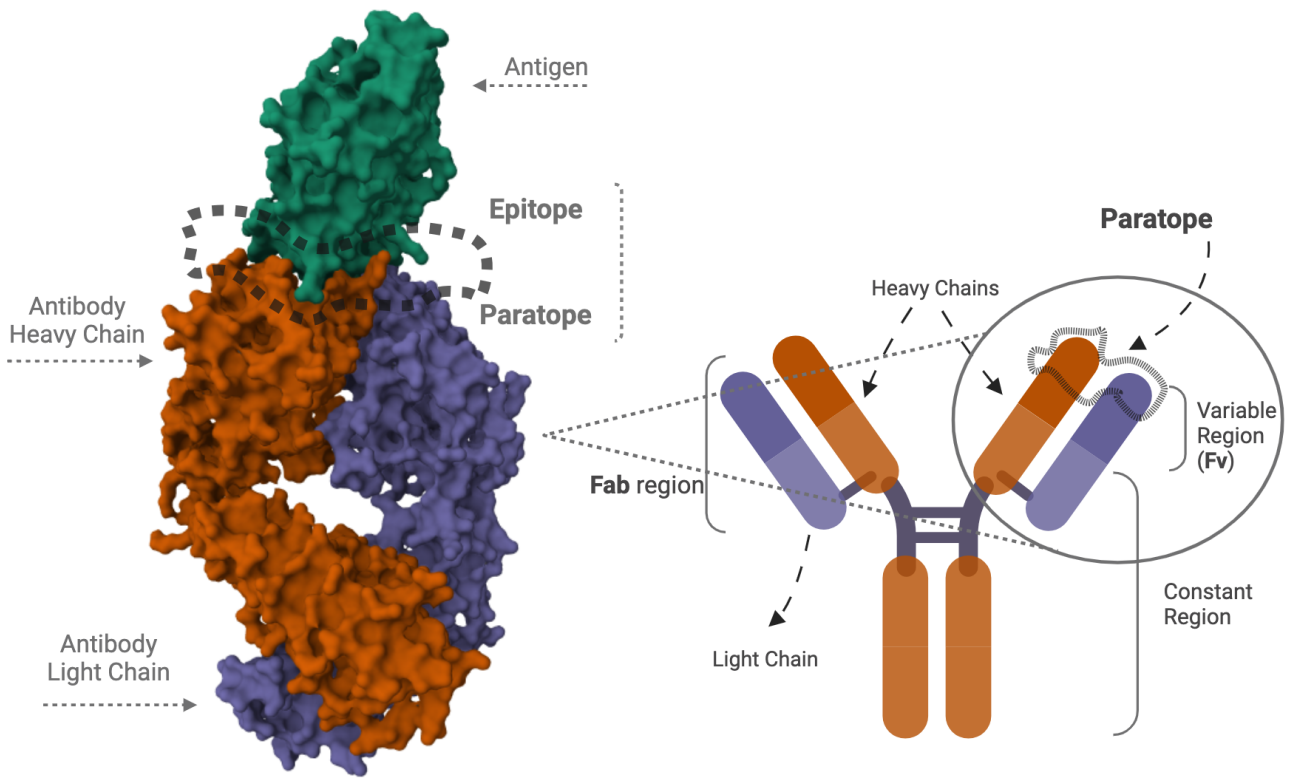

**Figure 5.** Representation of main Antibody regions (IgG). The secondary structure of the Fragment antigen-binding region (Fab region), on the left side. On the right side, a schematic portrayal of the Fab; the Heavy (dark and light red colour) and Light (dark and light purple colour) chains; Paratope region mainly within the variable region (Fv).

The majority of paratope predictors have trained their machine learning algorithms using data exclusively from the CDRs or Fv region. This choice is driven by the need to address the inherent imbalance between binder and non-binder residues in the antibody-antigen complex. However, there are contrasting viewpoints. Some researchers have argued that amino acid modifications occurring outside the CDRH3 also play a role in influencing binding [65], [66]. Additionally, it has been noted that around 20% of paratope residues may reside outside the complementary determining regions [67], [68]. This observation suggests that a more comprehensive approach involving other antibody segments would be more suitable for effective model training [69].

**Machine learning-based tools**

The past decade has seen the emergence of several machine learning-based frameworks for the task of paratope prediction. These frameworks include ParaPred (2018)[70], Daberdaku and Ferrari (2019)[71], PECAN (2020)[72], proABC-2 (2020)[73], DeepANIS (2021)[74], and Paragraph (2023)[75]. While ParaPred and proABC-2 offer user-friendly web-based platforms, the others have made their codes available on GitHub. These frameworks follow a similar workflow to the predictive tools discussed in the epitope sections, expecting as input an antibody sequence in FASTA format or its structure in the PDB format, to classify residues as either belonging to the paratope or not. Notably, PECAN also requires the antigen sequence as part of the input.

All of these frameworks are designed via supervised learning. ParaPred and proABC-2 employ Convolutional Neural Networks as their primary algorithms, with ParaPred additionally incorporating a standard Long Short-Term Memory (LSTM) architecture. DeepANIS combines bidirectional LSTM with Transformer encoder and Multilayer Perceptron (MLP). On the other hand, Daberdaku *et al.*, PECAN, and Paragraph opted for Support Vector Machine, Graph Attention Network (GAN), and Equivariant Graph Neural Network (EGNN), respectively.

**Datasets**

Structural databases containing antibody-antigen complexes were explored by these tools. proABC-2 sourced its data from the PDB, subsequently employing isotype-specific Hidden Markov Models (HMM) profiles to identify antibody-antigen complexes. On the other hand, the rest of the methods preferred sources containing antibodies that were already annotated. ParaPred, DeepANIS, and Paragraph utilised SAbDab, while Daberdaku et al. and PECAN made use of AbDb.

ParaPred and DeepANIS followed a similar approach by selecting 277 antibody-antigen complex structures and focusing on their CDRs regions. They extended these regions by adding four additional residues, two at each terminal. Daberdaku et al. merged datasets from ParaPred and two other prior tools [68], [76] creating a split of 213 complex structures for training and 106 for validation. An additional set of 153 structures was compiled for testing, drawn from AbDb. PECAN retained the datasets from Daberdaku *et al.* with minor adjustments. It considered the Fv regions from antibodies and selected only complexes, resulting in 205 structures for training and 152 for testing. In contrast, Paragraph initially employed the same dataset as PECAN but introduced an updated set of 651 structures for training, along with 217 for internal validation and 218 for testing, sourced from SabDAb. Similar to ParaPred, Paragraph employed the extended CDR region as its input data. proABC-2 curated a larger dataset of 769 Ab-Ag complexes from the PDB for training its model.

For locating the starting and ending positions of CDRs in antibody sequences, Parapred, DeepANIS, and proABC-2 adopted the Chothia numbering scheme [77], while PECAN and Paragraph opted for IMGT [30]. In contrast, Daberdaku *et al.* introduced a distinct approach based on a geometric representation, in which the 3D structure of the antibody is considered and its entire surface divided into small regions. Despite differences in the numbering, the criterion for defining and labelling paratope residues remained consistent across all tools, relying on a distance threshold of 4.5Å between any heavy atoms of the antigen.

Additionally, apart from crystal structures of antibodies, Paragraph and PECAN generated data of model structures using the ABodyBuilder platform [78], utilising the sequences from their curated datasets as references. This was done to assess the framework's capabilities using antibody models (Fv region only).

**Feature Engineering**

Different strategies were investigated jointly to encode antibody sequence and structure, as described in Table 6. The majority of tools converted the amino acid sequences into vectors using either one-hot encoding, as seen in Parapred, PECAN, proABC-2, and Paragraph, or integer vectors, the choice of DeepANIS.

Furthermore, Paragraph enhanced the vector by incorporating a binary representation indicating the antibody chain type, whether heavy or light. Drawing from amino acid representations outlined in [79], Parapred incorporated an additional seven numerical parameters capturing physicochemical and structural attributes. These parameters encompassed steric parameter, polarizability, volume, hydrophobicity, isoelectric point, helix probability, and sheet probability.

Evolutionary insights were harnessed by PECAN and DeepANIS through the utilisation of the PSSM and the predicted ASA. DeepANIS went a step further by combining the HMM profile derived from the UniProtKB database [80] with a 30% identity threshold (UniClust30). This integration enabled the prediction of secondary structure, backbone torsion angles, and half-sphere exposure. In the case of PECAN, the RSA was also leveraged along with iterative measurements of amino acid frequency within a 8Å radius.

proABC-2 took a different approach by employing binary vectors to represent arrays of aligned sequences for both heavy and light chains. This encompassed information about germline family, canonical structures, and hypervariable loop lengths. In contrast, Daberdaku *et al.* transformed portions of the antibody surface, considering specific distances, into spherical shapes. These shapes were then converted into vectors termed 3D Zernike descriptors (3DZDs), which were further enhanced with 20 antibody-related indices sourced from the AAIndex database.

Since Paragraph and PECAN are based on graph neural networks, their selected encoding approaches are on top of inherent graph embeddings.

Table 6. The collection of antibody encoding strategies exploited in each tool is detailed under "Features" in the second column, with the corresponding tool name in the first column.

| Tools | Features |
|---|---|
| ParaPred | AA one-hot-encoding, numerical indices (Physical, Chemical, and Structural) |
| Daberdaku et al. | 3D Zernike, AAindex |
| PECAN | AA one-hot-encoding, PSSM, ASA, RSA, amino acid frequency profile |
| proABC-2 | AA one-hot-encoding, canonical structures and length of hypervariable loops, germline family |
| DeepANIS | AA integer encoding, PSSM, ASA, HMM profile, secondary structure, backbone torsion angles and half-sphere exposure |
| Paragraph | AA and Chain type one-hot-encoding |

**Performance**

The tools described here assessed the performance of their machine learning models through either a cross-validation approach or a blind test using an independent dataset. Notably, Daberdaku *et al.*

uniquely combined both strategies to ensure a more robust evaluation. Table 7 provides an overview of the selected algorithm, validation technique, statistical metrics, training and testing dataset sizes, and results for each tool. A consistent trend of performance improvement is observed across these tools, despite occasional variations in the evaluation datasets. This trend aligns with the common data source. Notably, the most recent tool, Paragraph, demonstrated superior performance compared to PECAN, Parapred, and Daberdaku *et al.* using a shared dataset, as highlighted in its corresponding publication.

**Table 7.** A comparative analysis of Antibody Paratope predictors includes information on the algorithm used, validation approach (cross-validation, blind-test), the set of statistical metrics employed, dataset sizes, as well as model performance under cross-validation and on blind-test with the independent dataset.

| Tools | Algorithm | Validation | Metrics | Dataset Size | Results on CV | Results on Independent Set |
|---|---|---|---|---|---|---|
| ParaPred | CNN + LSTM | 10-fold CV using training set | MCC, F1, ROC-AUC | Training: 277 structures | MCC 0.554; F1 0.690; ROC-AUC 0.878 | N/A |
| Daberdaku et al. | SVM | 10-fold CV using training set; Blind-test with Independent set | ROC-AUC, PR-AUC | Training: 213 structures; Testing: 153 structures | ROC-AUC 0.895 | ROC-AUC 0.950; PR-AUC 0.658 |
| PECAN | GAN | Blind-test with Independent set | ROC-AUC, PR-AUC, Precision, Recall | Training: 205 structures; Testing: 152 structures | N/A | ROC-AUC 0.96; PR-AUC 0.70; Precision 0.42; Recall 0.922 |
| proABC-2 | CNN | 10-fold CV using training | ROC-AUC, MCC, F1 | Training: 769 | ROC-AUC 0.96; MCC | N/A |

| | | set | | structures | 0.57; F1 0.59 | |
|---|---|---|---|---|---|---|
| DeepANIS | BiLSTM + Transformer Encoder + MLP | 10-fold CV using training set | MCC, PR-AUC | Training: 277 structures | MCC 0.606; PR-AUC 0.727 | N/A |
| Paragraph | EGNN | Blind-test with Independent set | PR-AUC, ROC-AUC, F1, MCC | Training: 651 structures; Testing: 218 structures | N/A | PR-AUC 0.725; ROC-AUC 0.934; F1 0.696; MCC 0.669 |

**Significance and Limitations**

An improved understanding of the various possible conformations within the Antibody hypervariable regions [81], coupled with insights into evolutionary patterns and the diverse nature of CDRH3 loops [11], has significantly advanced antibody characterization and focused efforts on paratope identification.

The advancements witnessed in machine learning-based paratope predictors hold promising implications for antibody-related applications. These predictors could play a pivotal role in prioritising potential dy antibody designs. Moreover, architectural innovations like EGNN [82], capable of graphically embedding 3D inputs while maintaining equivariance to rotations, translations, reflections, and permutations, have progressively contributed to more accurate representations of complex biological entities like antibodies and dynamic organisational domains.

Similar to epitope predictors, the intricate interplay between antibodies and antigens, along with the diverse conformational possibilities arising from their interactions, remains a current challenge that most tools do not yet fully address. Furthermore, the dynamic nature of paratope regions, given their extensive sequence and structural diversity, presents an open question and a significant hurdle in capturing the complete essence of paratope regions from available data.

# ANTIBODY DESIGN

The design of an antibody capable of effectively exerting a desired biological function against a target epitope offers a promising alternative to traditional treatments for complex diseases. Major outcomes of the antibody design process are depicted in Figure 6, which encompass the generation of novel amino acid sequences for either the complete or partial regions (either the variable or the CDRs) and the subsequent modelling of the 3D structure, possessing enhanced properties and functions in a realistic context.

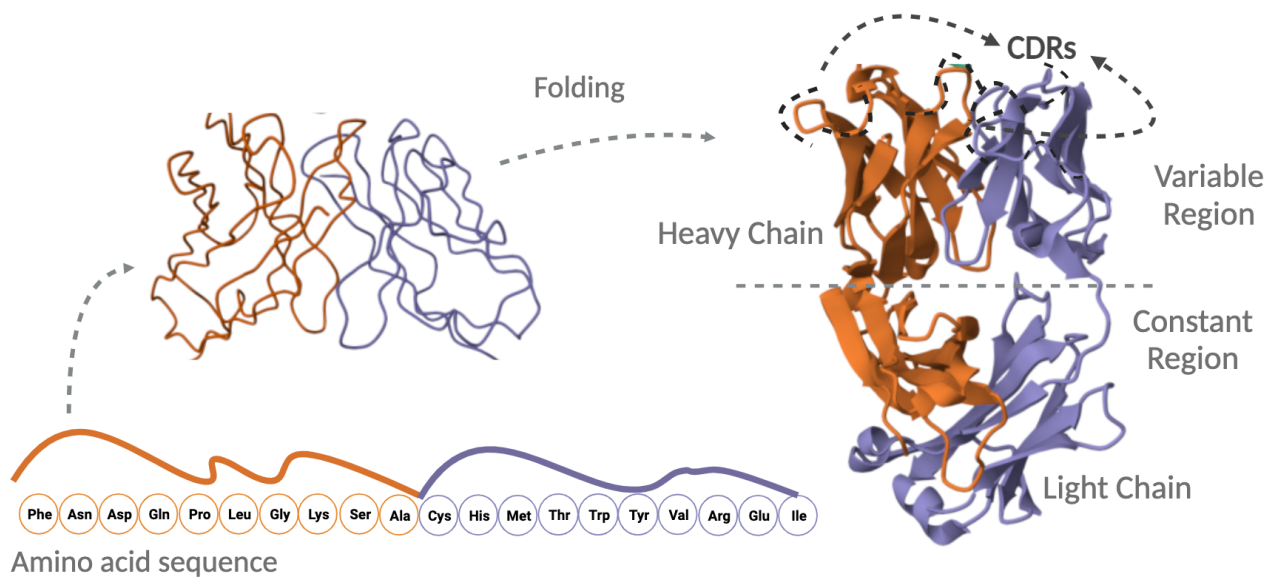

**Figure 6.** Antibody Design is represented as the creation of novel antibody sequences, on the bottom left side, and confident prediction of partially or full region structures, depicted as the Fab region on the right side.

The design process goes beyond the consideration of paratope-epitope binding affinity and involves overarching criterias, including developability. Developability encompasses a spectrum of chemical, physicochemical, and biophysical properties with the goal of generating antibodies that adhere to stringent drug-like standards, ensuring measurable potency, safety, and compatibility with pharmaceutical manufacturing. This has been extensively detailed in prior reviews [83], [84]. Furthermore, these reviews have comprehensively addressed the evolution of experimental and computer-aided procedures in Antibody design [85], the achievements through the combination of diverse *in silico* methods [86]–[88], and the specific enhancements within antibody design for oncology and organ delivery mechanisms, as highlighted in [89].

The Therapeutic Antibody Profiler (TAP) [90] introduced guidelines based on the analysis of properties from 242 clinical trials involving antibody therapeutics that had undergone phase 1. These guidelines are distilled into five key characteristics that indicate weak developability: the amino acid length of CDRs, the extent of hydrophobicity across the surface, the charge within the CDRs (both positive and negative), and the net charge imbalance between the heavy and light chain surfaces. Additionally, they noted that not all antibodies derived from humans are inherently favourable therapeutic agents.

In contrast, for assessing the human-likeness of antibodies, AbDiver [91] was developed as a platform that compares a given sequence with a repertoire of naturally observed antibodies, obtained through NGS techniques. Additionally, BioPhi [92] applied a deep-learning transformer model to identify non-human amino acid stretches in antibody sequences, recommending targeted substitutions. Similarly, Hu-mAb [93] utilised random forest classifiers to detect non-human sequences within the variable domain proposing mutations towards humanization.

Leipold and Prabhu [94] explored the benefits and ongoing challenges of optimising pharmacokinetic and pharmacodynamic properties of antibodies, highlighting how improved understanding and prediction of both aspects contribute to advancements in antibody design.

Thus, the primary focus of this section is on frameworks specifically dedicated to *de novo* antibody generation, which refers to machine learning-based methods aimed at generating new realistic antibody sequences or structures. This section is organised as follows: a description of Generative Models and their substantial contributions in this field, a presentation of Antibody Design tools categorised according to their defined objectives, a summary of data and evaluation metrics, and an exploration of their Significance and Limitations.

*Generative Models*

In contrast to discriminative models, generative models are a class of statistical models that can generate authentic and novel instances by utilising an inferred probability distribution derived from the underlying distribution of the training data [95]. In this context, generative models aim to capture the joint probability of independent and dependent variables in supervised scenarios or the marginal probability of the independent variable in unsupervised scenarios.

This class of machine learning has been applied to the generation of new antibody sequences, and various types of Deep Neural Networks (DNN) have been explored for this purpose, reflecting trends in protein design more broadly [96]. Examples include Transformer-based Language Models, Hierarchical Message Passing Network (HMPN), Long Short-Term Memory (LSTM), Generative Adversarial Network (GAN), Variational Auto-Encoder (VAE), and Graph Neural Network (GNN).

**Antibody Design Tools**

The majority of the existing literature on this subject consists of recent preprints. While these tools provide methodological insights and findings, only a few have been made openly available through standalone packages or by sharing their source code in online repositories. Most of these tools have focused on designing antibodies for specific regions. For instance, Amimeur *et al.* [97] and Saka *et al.* [98] concentrated on sequence generation for the Variable region (Fv), whereas DiffAb [99], Gao *et al.* [100], Jin *et al.* [101], MEAN [102], and dyMEAN [103] centred their efforts on sequence and structure modelling for the complementary determining regions. In a broader context, IgML [104] introduced full sequence design for both heavy and light chains. Consequently, each of these tools implemented distinct frameworks with specific goals, ranging from expanding human and multispecies repertoire of sequences, enhancing binding affinity toward a particular antigen, and generating new sequences and structures exclusively for the CDR.

*Human Repertoire sequences*

With the aim of creating new sequences to broaden the human repertoire while maintaining assessed biophysical characteristics, Amimeur *et al.* developed a framework based on a GAN architecture. The objective was to generate variable domains of light and/or heavy chains that are similar to those found in humans, with a length of 148 residues. The Generator component of the architecture was initialised using a noise vector, while the Discriminator (classifier) underwent progressive training using both genuine human antibody sequences and the outputs created by the generator.

The validation of artificially generated sequences was carried out experimentally. A subset of 100,000 sequences, resulting from the combination of four germline sequences, was expressed using phage display technique. Subsequently, these sequences were analysed to confirm beneficial biophysical properties.

*Multispecies Antibody library*

In an effort to expand antibody libraries across various species while improving developability attributes, the Immunoglobulin Language Model (IgLM) employed a 4-layer GPT-2 Transformer architecture. Previously optimised unpaired sequences were sourced from diverse antibody repertoires to serve as a source for generating novel sequences based on factors such as chain type (heavy or light) and species (human, mouse, camel, rat, rabbit, or rhesus).

For the training process, specific sections of the input sequences were masked, and then filled in with content influenced by their surrounding context. To assess whether the newly generated sequences possessed the desired developability attributes, external computational tools were employed. The

OASis score [92] was used to evaluate humanness, while the SAP score [105] and CamSol were [106] used for aggregation and solubility predictions, respectively.

Focusing on enhancing binding affinity toward the specific kynurenine antigen, Saka and colleagues designed a 2-layer LSTM architecture based on optimised anti-kynurenine antibody sequences. The intention was for the model to learn the underlying properties of these effective binders during training and then use that knowledge to generate new amino acid residues character by character.

To validate the results, an experimental step was taken involving some of the generated samples. The Dissociation constants were assessed using surface plasmon resonance, revealing an expected positive correlation between the model's probability distribution and the binding affinities.

*CDRs Generation: sequence and structure*

DiffAb, Gao *et al.*, and Jin *et al.* developed methods that require prior knowledge of the antibody framework region as input. These methods then proceed to design both the sequence and structure of the CDRs. An additional requirement for DiffAb is the input of a bound antigen structure. In this case, DiffAb employs two Markov chains (a forward and a generative diffusion process) along with the Rosetta side-chain packing tool [107] to generate the CDR structure. Gao *et al.* also requires the epitope structure in complex with the antibody as input. They employed a pre-trained antibody language model (AbBERT) to initialise the residues instead of using random initialization. Their approach involved two Hierarchical Message Passing Networks for sequence generation and structure prediction. While demanding the antigen of interest might potentially lead to more customised CDRs, it also comes with limitations due to the requirement of prior knowledge about the Ab-Ag complex and their relative orientation.

Jin *et al.* introduced the RefineGNN, a graph neural network for generating graph representations specifically for the heavy chain CDRs. This model was combined with a Message Passing Network (MPN) and a Recurrent Neural Network (RNN) with attention mechanisms. This approach helps model residue nodes in blocks, which reduces the length of sequences and enhances context propagation through graph convolutions.

MEAN and DyMEAN utilised equivariant graph neural networks and included multi-channel inputs to represent the different atoms constituting a residue. MEAN focused on backbone atoms, while DyMEAN further considered the variability of side chains. DyMEAN used the epitope structure and the antibody sequence as input (except the CDRH3), generating the entire antibody 3D structure. On the other hand, MEAN required the bound antibody-antigen structure as input, omitting the CDRs of the heavy chain, and proceeded to design both the sequence and structure of all three CDRs in the heavy chain.

**Summary**

Table 8 provides a summary of the tools, their corresponding machine learning architecture, the target antibody region, source database, the type of evaluation, and their potential for reuse. The tools discussed in this context primarily leveraged the OAS repository and the SAbDab database to source diverse antibody sequences and structures for their training purposes. Notably, Saka *et al.* uniquely utilised an in-house biopanning approach via phage display, focusing on the specific F02 antibody (anti-kynurenine) to generate a specific dataset.

The evaluation strategies employed can be categorised into experimental and computational approaches. Among these tools, MEAN was the only one to mention using a 10-fold cross-validation technique to assess the generalisation of their models. Both Amimeur *et al.* and Saka *et al.* carried out post-hoc experimental validations using the generated antibody sequences. Furthermore, Saka *et al.* detailed the application of the negative logarithm likelihood (NLL) as a metric and highlighted its use in prioritising the predicted sequences, suggesting that lower NLL values correspond to higher binding affinity toward the specific antigen.

Other tools employed various computational metrics to define success, predominantly including Perplexity and Amino Acid Recovery Rate (AAR), which are associated with sequence generation, as well as Root Mean Square Deviation (RMSD), which pertains to predicted structure. Perplexity, a frequently used metric in Language Models, provides a probability score indicating the model's uncertainty when generating a new sentence (lower values are preferable), in this context, a new antibody sequence. AAR measures the percentage similarity between a newly generated sequence and the ground truth, aiming for values closer to 100% for higher identity.

For structure-based assessments, RMSD, the Template Modelling Score (TM-score) [108] and the Local Distance Difference Test (IDDT) [109] offer measures of global and local similarity between two structures, respectively.

The Evaluation column in Table 8 demonstrates the performance outcomes of the presented tools based on these computational metrics, as reported in their original publications, standardised to the CDR region for better comparison. Except for IgLM, the remaining tool performances are based on the same benchmark dataset, extracted from the RAbD [110][90], consisting of 60 antibody-antigen structures.

In generating new sequences and structures for the third CDR in the heavy chain (CDRH3), most tools produced RMSD values below 3Å, demonstrating strong alignment with actual structures. Given that protein function depends on its folded structure, these results are highly meaningful, especially as function can be preserved even with just 30% sequence identity in the new structures.

**Table 8.** List of machine learning based tools trained to design antibodies. The Architecture column refers to the main algorithm implemented; the Target Region Design column indicates the chosen area of the Antibody that is being created; the Data Source shows the origin of the training dataset; the amount of sequences or structures is displayed in the Training dataset column; the Evaluation column briefs whether an Experimental or Computational (the metrics) evaluation was conducted; the last column indicates if the corresponding codes were made available.

| Tools | Architecture | Target Region Design | Data Source | Training dataset | Evaluation | Tool Usability |
|---|---|---|---|---|---|---|
| Amimeur et al. | GAN | Fv region | Public: Observed Antibody Space (OAS) | 400,000 sequences | Experimental | N/A |
| Saka et al. | LSTM | Fv region heavy chain | In-house: Phage Display Biopanning | 959 sequences of F02 heavy chain | Experimental | N/A |
| IgLM | GPT-2 Transformer | Heavy and Light chain sequences | Public: Observed Antibody Space (OAS) | 558M light chain variable sequences | **Perplexity**: CDRH1/CDRH2=1.5; CDRH3=4.5 | GitHub |
| DiffAb | Diffusion-based Markov chains | CDR | Public: SAbDab | Not informed | For CDRH3: **RMSD**=3.597 Å **AAR**=26.78% | GitHub |
| Gao et al. | AbBERT + HMPN | CDR Heavy chain | Public: OAS + SAbDab | Pre-training: 50M sequences; 11,822 CDR sequences heavy chain | For CDRH3: **RMSD**=1.62Å **AAR**=40.35% | N/A |
| Jin et al. | GNN + MPN | CDR Heavy chain | Public: SAbDab | 11,822 CDR sequences heavy chain | For CDRH3: **RMSD**=2.50Å **AAR**=35.37% | N/A |
| MEAN | EGNN | CDR Heavy chain | Public: SAbDab | 3,127 antibody structures | For CDRH3: **RMSD**=1.81Å **AAR**=36.77% **TM-score**=0.98 | GitHub |

| dyMEAN | EGNN | CDRH3 | Public: SAbDab | 3,256 antibody structures | For CDRH3: **AAR**=43.65% **TM-score**=0.97 **LDDT**=0.84 | GitHub |

**Significance and Limitations**

The potential of generative models to advance antibody design has been increasingly recognised across various studies and applications. These models have been instrumental in producing sequences and structures with desired attributes, enabling them to uncover underlying patterns and connections within vast datasets. Traditional experimental antibody screening and optimisation processes are resource-intensive and often fall short of guaranteeing optimal outcomes. In contrast, *de novo* antibody generation offers a more controlled and scalable approach, capable of exploring and capturing the intricate relationships among sequence, structure, and function.

The tools presented in this section have devised diverse workflows to harness generative models, along with evaluating their performance. These tools aim to create novel antibody sequences and structures, ranging from broad human repertoires to specific antibody-antigen complexes. Some tools had the advantage of a suitable lab setup to experimentally validate a portion of their results, assessing improvements in antigen binding affinity as well as immunogenicity and diversity, which are naturally present in human antibodies.

While many tools quantitatively assess new instances using metrics like Amino Acid Recovery Rate and Root Mean Square Deviation to measure sequence and structure similarity between generated and original samples, these metrics alone are insufficient to capture all critical aspects of antibody design. Key factors such as functional properties, binding affinity, and epitope coverage are not fully addressed by AAR and RMSD. Moreover, these metrics rely solely on known antibodies for comparison.

Hence, evaluating the performance of generative models in antibody design, as an application-specific task, would benefit from incorporating additional tools. These could include *in silico* prediction methods for biophysical properties, epitope-paratope interactions, binding and developability affinity. Experimental validation remains vital to ensure that the generated antibodies possess desired characteristics beyond mere sequence and structure. Furthermore, the current lack of standardised comprehensive assessment in this field is an area that requires attention.

**CONCLUSIONS**

Immunoinformatics has witnessed significant evolution in the last decade due to advanced protein modelling and engineering techniques. However, a comprehensive understanding of epitope-paratope interactions and their dynamics remains unclear. This challenge is particularly compounded by the vast diversity in antibody sequences, the wide range of potential antigens and their dynamic nature. Novel problem formulations, enhanced mathematical abstractions, and improved modelling approaches are required to address this complexity.

In this context, we have identified and discussed machine learning-based tools that have emerged over the past decade for predicting Linear and Conformational B-cell Epitopes, as well as Paratopes. These tools, accessible as web servers or through online repositories, have developed frameworks that take antibody or antigen data as input, encompassing steps such as data encoding, feature generation, and classification using supervised machine learning models.

Notable enhancements have been observed in model performance, encoding methods, statistical analysis, and output visualisation (in the case of web servers). Despite their significant value, these tools are not yet optimal when used in isolation for predicting epitopes and paratopes in practical applications. Both fields could benefit from standardised benchmark datasets for evaluating progress and the adequate use of statistical metrics that provide equitable evaluation of predictions.

Furthermore, we have explored the application of machine learning architectures in Antibody design, which is one of the potential applications stemming from these prediction capabilities. Generative Models are playing a pivotal role in this area, either by creating new antibody sequences or repertoires, or by enhancing existing ones to achieve improved binding affinity between paratopes and epitopes.

The need for additional experimental settings to validate the efficacy of computational models is crucial for translating these research advancements into biopharmaceuticals. This could pave the way for the development of next-generation vaccines and immunotherapies.

**FUNDING**

This work was supported by Melbourne Research Scholarship; an Investigator Grant from the National Health and Medical Research Council of Australia [GNT1174405]. Supported in part by the Victorian Government's OIS Program.

**CONFLICT OF INTEREST**

The authors declare no conflict of interest.